\documentclass[11pt,a4paper]{article}

\usepackage{graphicx}
\usepackage{amssymb}

\usepackage{times}
\usepackage[monochrome]{color} 
\usepackage[normalem]{ulem}

\begin{document}

\newcommand{\be}{\begin{equation}}
\newcommand{\ee}{\end{equation}}
\newcommand{\bea}{\begin{eqnarray}}
\newcommand{\eea}{\end{eqnarray}}
\newcommand{\f}{\frac}
\newcommand{\p}{\partial}
\newcommand{\no}{\nonumber}
\newcommand{\kT}{k_{\rm B}T}
\newcommand{\e}{{\rm e}}
\newcommand{\dd}{{\rm d}}

\newcommand{\eg}{\textit{e.g.}}
\newcommand{\ie}{\textit{i.e.}}
\newcommand{\etc}{\textit{etc}}

\newcommand{\comdeg}{d^{\rm com}}
\newcommand{\ovsize}{s^{\rm ov}}
\newcommand{\comsize}{s^{\rm com}}

\newcommand{\kav}{\left<k\right>}

\newcommand{\corr} {\color{red}}
\newcommand{\tamas}{\color{blue}}
\newcommand{\imre} {\color{green}}
\newcommand{\illes}{\color{magenta}}
\newcommand{\gergo}{\color{cyan}}
\newcommand{\misc} {\color[rgb]{.62,.32,.18}}   

\begin{Large}
\begin{center}
\textbf{
Uncovering the overlapping\\
community structure of complex\\
networks in nature and society
}
\end{center}
\end{Large}

\medskip

\begin{large}
\begin{center}
Gergely Palla$^{\dagger\ddagger}$,
Imre Der\'enyi$^{\ddagger}$,
Ill\'es Farkas$^{\dagger}$, and
Tam\'as Vicsek$^{\dagger\ddagger}$
\end{center}
\end{large}

\bigskip

$^{\dagger}$Biological Physics Research Group of HAS,
 P\'azm\'any P.\ stny.\ 1A, H-1117 Budapest, Hungary,

$^{\ddagger}$Dept. of Biological Physics, E\"otv\"os University,
 P\'azm\'any P.\ stny.\ 1A, H-1117 Budapest, Hungary.


\baselineskip=0.55truecm
\bigskip
\bigskip

\textbf{
Many complex systems in nature and society can be described in terms
of networks capturing the intricate web of connections among the units
they are made of
\cite{watts-strogatz,barabasi-albert,albert-revmod,dm-book}.
A question of interest is how to interpret
the global organisation of such networks as the coexistence of their
structural sub-units (communities) associated with
more highly interconnected parts.  Identifying these \textit{a priori}
unknown building blocks (functionally related  proteins
\cite{ravasz-science,spirin-pnas},
industrial sectors
\cite{onnela-taxonomy},
groups of people
\cite{scott-book,watts-dodds},
\etc.) is crucial to
the understanding of the structural and functional
properties of networks. The existing
deterministic methods used for \emph{large} networks find separated
communities, while most of the actual networks are made of highly
overlapping cohesive groups of nodes. Here we introduce an approach to
analyse the main statistical features of the interwoven
sets of overlapping communities 
making a step towards the uncovering
of the modular structure of complex systems.
After defining
a set of new characteristic quantities for the statistics of communities,
we
apply an efficient technique to explore overlapping communities on a
large scale. We find that overlaps are significant, and
the distributions we introduce reveal universal
features of networks. Our studies of collaboration, word association,
and protein interaction graphs demonstrate that the \emph{web of
communities} has non-trivial correlations and specific scaling
properties.
}

\begin{figure}[!t]
\centerline{\includegraphics[width=0.5\textwidth]{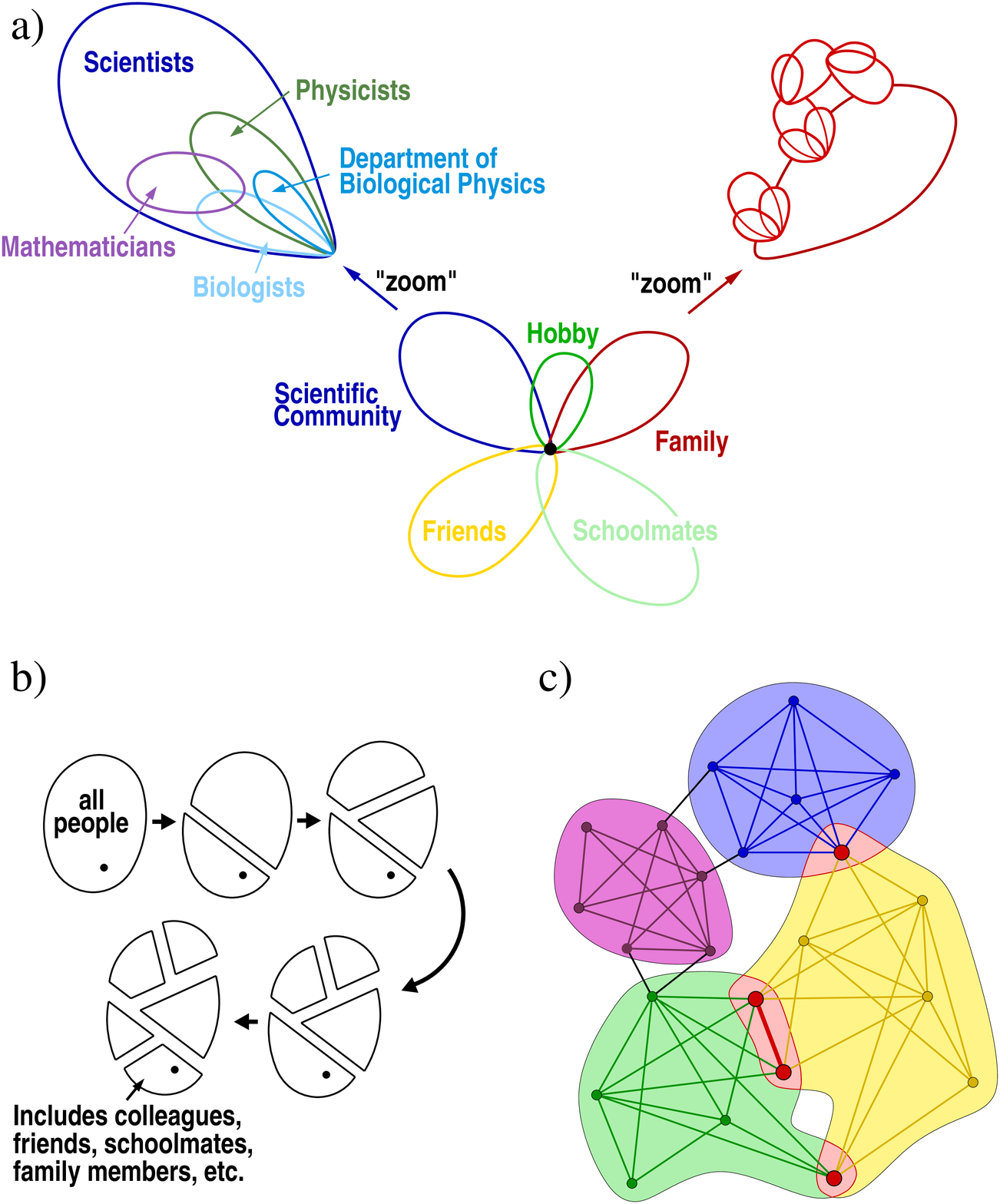}}
\caption{
Illustration of the concept of overlapping communities.
a) The black dot in the middle represents either of the authors of this
Letter, with several of his communities around. Zooming into the
scientific community demonstrates the nested and overlapping structure
of the communities, while depicting the cascades of communities
starting from some members exemplifies the interwoven structure
of the network of communities.
b)
Divisive and agglomerative methods
grossly fail to identify
the communities when overlaps are significant.
c) An example of overlapping $k$-clique-communities at $k=4$. The
yellow community overlaps with the blue one in a single node, whereas
it shares two nodes and a link with the green one. These overlapping
regions are emphasised in red. Notice that any $k$-clique (complete
subgraph of size $k$) can be reached only from the $k$-cliques of the
same community through a series of adjacent $k$-cliques. Two
$k$-cliques are adjacent if they share $k-1$ nodes.
}
\label{fig:szeml}
\end{figure}

Most real
networks typically contain parts in which the nodes
(units) are more highly connected to each other than
to the rest of the network. The sets of such nodes are usually called
clusters, communities, cohesive groups, or modules
\cite{scott-book,pnas-suppl,everitt-book,knudsen-book,newman-europhys},
having no widely accepted, unique definition.
In spite of this ambiguity
the presence of communities in networks is a
signature of the hierarchical nature of complex systems
\cite{ravasz-science,vicsek-nature}.
The existing methods for finding communities in \emph{large}
networks are useful if the community structure is such that it can be
interpreted in terms of separated sets of communities (see
Fig.~\ref{fig:szeml}b and Refs.
\cite{pnas-suppl,domany-prl,gn-pnas,radicchi-pnas,newman-pre}).
However, most real networks are
characterised by well defined statistics of overlapping and nested
communities. Such a statement can be demonstrated by the
numerous communities each of us belongs to, including those related to
our scientific activities
or personal life (school, hobby, family) and so on, as illustrated in
Fig.~\ref{fig:szeml}a. Furthermore, members of our communities have
their own communities, resulting in an extremely complicated web of the
communities themselves. This has long been appreciated by sociologists
\cite{wasserman},
but has never been studied systematically for large networks.
Another, biological example is that a large fraction of proteins belong
to several protein complexes simultaneously
\cite{protein-complexes}.

In general, each node $i$ of a network can be characterised by a
\emph{membership number} $m_i$, which is the number of communities the
node belongs to. In turn, any two communities $\alpha$ and $\beta$
can share $\ovsize_{\alpha,\beta}$ nodes, which we define as the
\emph{overlap size} between these communities. Naturally, the
communities also constitute a \emph{network}
with the overlaps being their links.
The number of such links of community $\alpha$
can be called as
its \emph{community degree}, $\comdeg_{\alpha}$.
Finally, the size $\comsize_\alpha$ of any community $\alpha$ can most
naturally be defined as the number of its nodes.
To characterise the community structure of a large network we introduce
the \emph{distributions} of these four basic quantities. In
particular, we will focus on their \emph{cumulative distribution
functions} denoted by
$P(\comsize)$, $P(\comdeg)$, $P(\ovsize)$, and $P(m)$, respectively.
For the overlap size, \eg, $P(\ovsize)$ means the proportion of
those overlaps that are larger than $\ovsize$.
Further relevant statistical features will be introduced later.

The basic observation on which our community definition relies is that
a typical community consists of several complete (fully connected)
subgraphs that tend to share many of their nodes.
Thus, we define a community, or more precisely, a
\emph{$k$-clique-community} as a union of all \emph{$k$-cliques}
(complete subgraphs of size $k$) that can be reached from each other
through a series of adjacent $k$-cliques (where adjacency means sharing
$k-1$ nodes)
\cite{everett-connections,short-cycles,clique-percolation}.
This definition is aimed at representing the fact that it is an
essential feature of a community that its members can be reached
through well connected subsets of nodes. There are other parts of the
whole network that are not reachable from a particular
$k$-clique, but they potentially contain further
$k$-clique-communities. In turn, a single node can belong to several
communities.
All these can be explored systematically and can
result in a large number of overlapping communities
(see Fig.~\ref{fig:szeml}c for illustration).
{\corr
Note that in most cases relaxing this definition (\eg, by allowing
incomplete $k$-cliques) is practically equivalent to lowering the value
of $k$.}
{\corr
For finding meaningful communities, the way they are 
identified is expected to satisfy several
basic requirements:
it cannot be too restrictive,
%
should be based on the density of links,
is required to be local,
should not
yield any cut-node or cut-link
(whose removal would disjoin the community)
and, of course, should allow overlaps.
We employ the
community definition specified above,
because none of the others in the literature
satisfy all these requirements simultaneously
\cite{everett-connections,kosub}.
}

{\corr
Although the numerical determination of the full set of
$k$-clique-communities is a polynomial problem, 
we use an algorithm which is exponential, because it
is significantly more efficient for the
graphs corresponding to actual data. This
}
method is based on first locating
all cliques (maximal complete subgraphs) of the network and then
identifying the communities by carrying out a
standard component analysis of the
clique-clique overlap matrix
\cite{everett-connections}.
For more details about the method and its speed see the Supplementary
Information.

\begin{figure}[!t]
\centerline{\includegraphics[width=\textwidth]{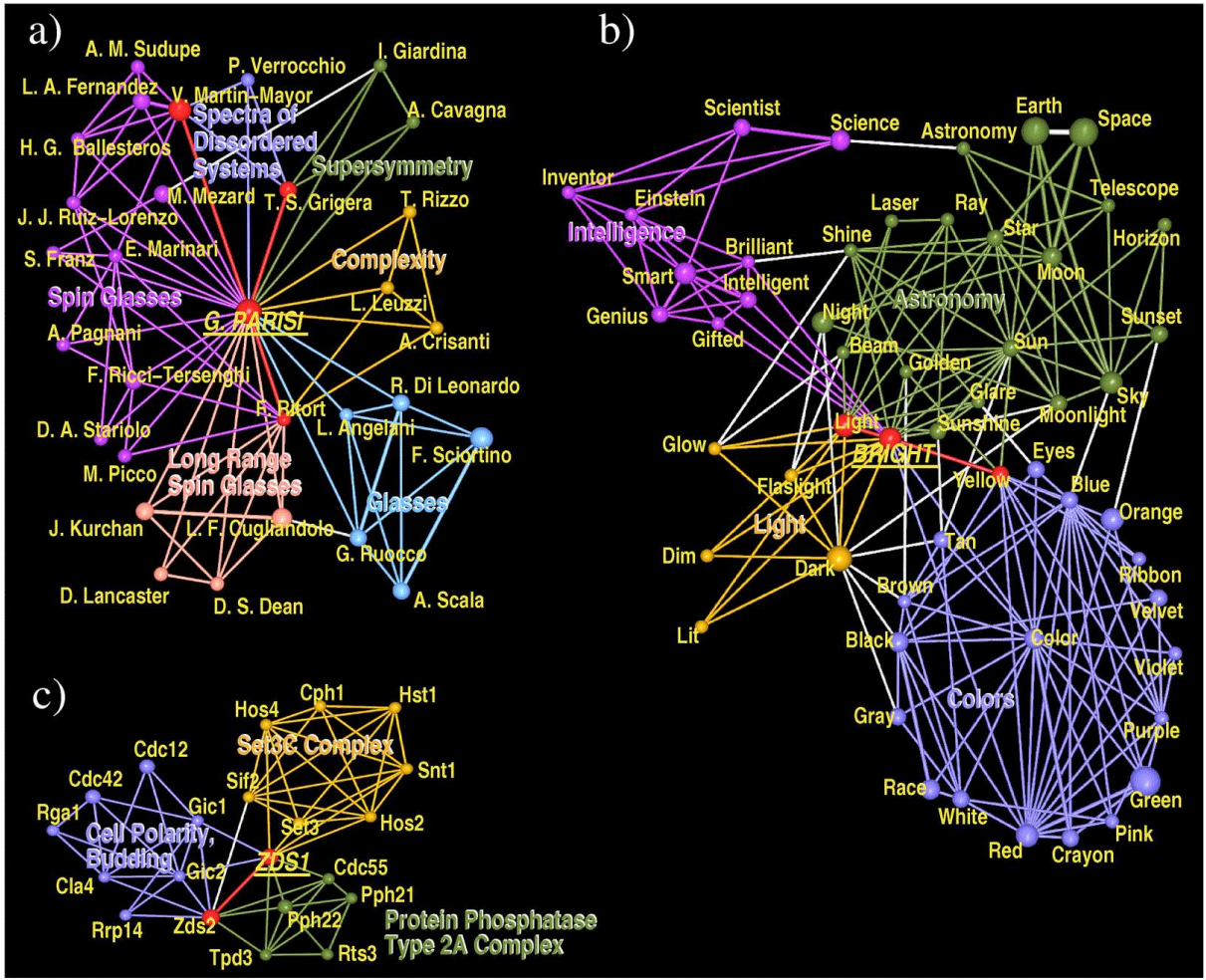}}
\caption{
The community structure around a particular node in three
different networks.
The communities are colour coded, the overlapping
nodes and links between them are emphasised in red, and the volume of
the balls and the width of the links are proportional to the total
number of communities they belong to. For each network the value of $k$
has been set to 4.
a) The communities of G. Parisi in the co-authorship network
of the Los Alamos cond-mat archive (for threshold weight $w^*=0.75$)
can be associated with his fields of interest.
b) The communities of the word ``bright'' in the South Florida Free
Association norms list (for $w^*=0.025$) represent the different
meanings of this word.
c) The communities of the protein ZDS1 in the DIP core list of the
protein-protein interactions of \textit{S. cerevisiae} can be
associated with either protein complexes or certain functions.
}
\label{fig:glay}
\end{figure}

We use our method for binary networks (\ie, with undirected
and unweighted links). An arbitrary network can always be
transformed into a binary one by
ignoring any directionality in the links and
keeping only those that are stronger than
a threshold weight $w^*$.
Changing the threshold is like changing the resolution (as in a
microscope) with which the community structure is investigated: by
increasing $w^*$ the communities start to shrink and fall apart. A very
similar effect can be observed by changing the value of $k$ as well:
increasing $k$ makes the communities smaller and more disintegrated, but
at the same time, also more cohesive.

\begin{figure}[!t]
\centerline{\includegraphics[width=0.5\textwidth]{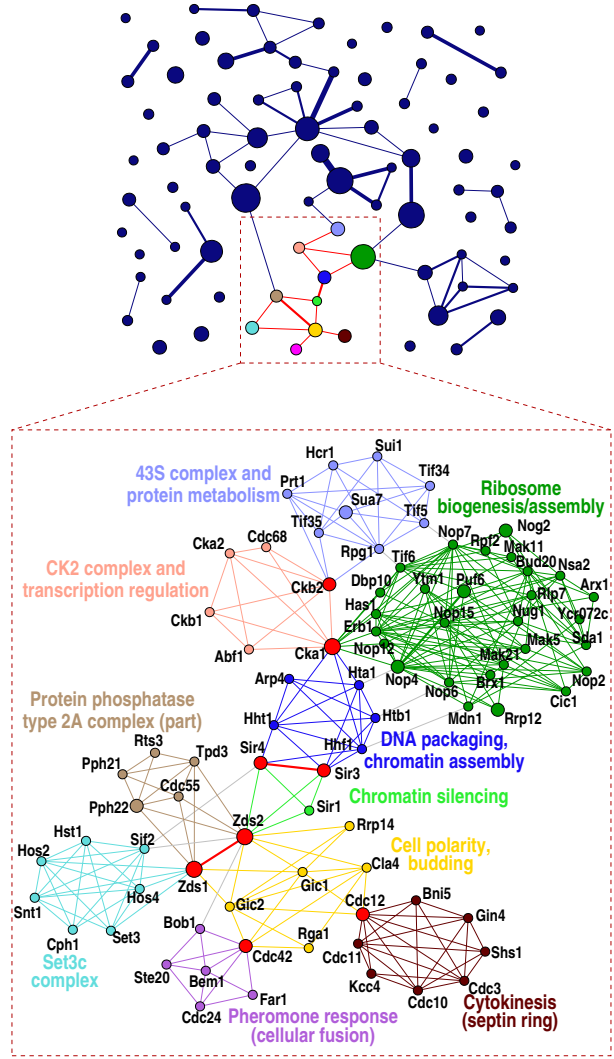}}
\caption{
Network of the 82 communities in the DIP core list of the
protein-protein interactions of \textit{S. cerevisiae} for $k=4$.
The
area of the circles and the width of the links are proportional to the
size of the corresponding communities ($\comsize_\alpha$) and to the
size of the overlaps ($\ovsize_{\alpha,\beta}$), respectively. The
coloured communities are cut out and magnified to reveal their internal
structure. In this magnified picture the nodes and links of the
original network have the same colour as their communities, those that
are shared by more than one community are emphasised in red, and the
grey links are not part of these communities. The area of the circles
and the width of the links are proportional to the total number of
communities they belong to.
}
\label{fig:prot-comm}
\end{figure}

When we are interested in the community structure around a
particular node, it is advisable to scan through some ranges of
$k$ and $w^*$, and monitor how its communities change.
As an illustration, in Fig.~\ref{fig:glay} we are depicting
the communities of three selected nodes of three large networks:
(i) the social network of scientific collaborators
\cite{cond-mat},
(ii) the network of word associations
\cite{nelson}
related to cognitive sciences, and
(iii) the molecular-biological network of protein-protein interactions
\cite{dip}.
These pictures can serve as tests or validations of the efficiency of
our algorithm.
In particular, the communities of the
author G. Parisi (who is well known to have been making significant
contributions in different fields of physics) shown in
Fig.~\ref{fig:glay}a are associated with his fields of interest, as it can
be deduced from the titles of the papers involved.
The 4-clique-communities of the word ``bright'' (Fig.~\ref{fig:glay}b)
correspond to the various meanings of this word. 
An important biological
application is finding the communities of proteins, based on their
interactions.
Indeed, most proteins in the communities shown in
Figs.~\ref{fig:glay}c and \ref{fig:prot-comm}
can be associated with either protein complexes or certain functions,
as can be looked up by using the GO-TermFinder package
\cite{termfinder}
and the online tools of the Saccharomyces Genome Database (SGD)
\cite{sgd}.
For some proteins no function is available
yet. Thus, the fact that they show up in our approach as members of
communities can be interpreted as a prediction for their functions.
{\corr
One such example can be seen in the enlarged portion of
Fig.~\ref{fig:prot-comm}.
For the protein Ycr072c, which is required for the viability of the
cell and appears in the dark green community on the right,
SGD provides no biological process (function).
By far the most
significant GO term for the biological process of this community is
``ribosome biogenesis/assembly''. Thus, we can infer that Ycr072c
is likely to be involved in this process.}
Also, new cellular processes can be predicted
if yet unknown communities are found with our method.

{\corr
These examples (and further ones included in the 
Supplementary Information) clearly
demonstrate the advantages of our approach over the existing
divisive and agglomerative methods recently used for large real networks.
Divisive methods cut the network
into smaller and smaller pieces, each node is forced to remain
in only one community and be separated from its other communities,
most of which then necessarily fall apart and disappear.
This happens, \eg, with the word ``bright'' when we apply
the method described in Ref.\
\cite{gn-pnas}:
it tends to stay together mostly with the words of the community
related to ``light'', while most of its other communities
(\eg, those related to ``colors'', see Fig.~\ref{fig:glay}b)
completely disintegrate
(``green'' gets to the vegetables, ``orange'' to the fruits, \etc.).
Agglomerative methods do the same, but in the reverse direction.
For example, when we applied the agglomerative method of Ref.\
\cite{newman-pre},
at some point ``bright'', as a single word, joined a 
``community'' of 890 other words.
In addition, such methods inevitably lead to a tree-like hierarchical
rendering of the communities, while our approach allows the
construction of an unconstrained network of communities.}

The networks chosen above have been constructed in the following ways.
In the co-authorship network of the Los Alamos e-print archives
\cite {cond-mat}
each article contributes the value $1/(n-1)$ to the weight of the
link between every pair of its $n$ authors.
In the South Florida Free Association norms list
\cite{nelson}
the weight of a directed link from one word to another indicates
the frequency that the people in the survey associated the end point of
the link with its start point. For our purposes these directed
links have been replaced by undirected ones with a weight equal
to the sum of the weights of the corresponding two oppositely directed
links.
In the DIP core
list of the protein-protein interactions of \textit{S. cerevisiae}
\cite{dip}
each interaction represents an unweighted link between the
interacting proteins.
These networks are very large, consisting of
 30739, 10617, and 2609 nodes and
136065, 63788, and 6355 links, respectively.

Although different values of $k$ and $w^*$ might be optimal for the
local community structure around different nodes, we should set some
global criterion to fix their values if we want to analyse the
statistical properties of the community structure of the entire
network. The criterion we use is based on finding a community
structure as highly structured as possible. In the related
\emph{percolation} phenomena
\cite{clique-percolation}
a giant component appears when the number of links is
increased above some critical point. Therefore,
to approach this critical point from below,
for each selected value of
$k$ (typically between 3 and 6) we lower the threshold $w^*$ 
until the largest community becomes
twice as big as the second largest one. In this way we ensure that we
find as many communities as possible, without the negative effect of
having a giant community that would smear out the details of the
community structure by merging many smaller
communities. We denote by $f^*$ the fraction of links stronger than $w^*$,
and use only those values of $k$ for which $f^*$ is not too
small (not smaller than 0.5). This has led us to $k=6$ and 5 with
$f^*=0.93$ and 0.75, respectively, for the collaboration network, and
$k=4$ with $f^*=0.67$ for the word association network.
For the former network both
sets of parameters result in very similar communities
(see the Supplementary Information). Since for
unweighted networks no threshold weight can be set, for these we
simply select the smallest value of $k$ for which no giant community
appears.  In case of the protein interaction
network this gives $k=4$, resulting in 82 communities.
Due to this relatively low number,
we can depict the entire network of protein communities
as presented in Fig.~\ref{fig:prot-comm}.

\begin{figure}[!t]
\centerline{\includegraphics[width=0.5\textwidth]{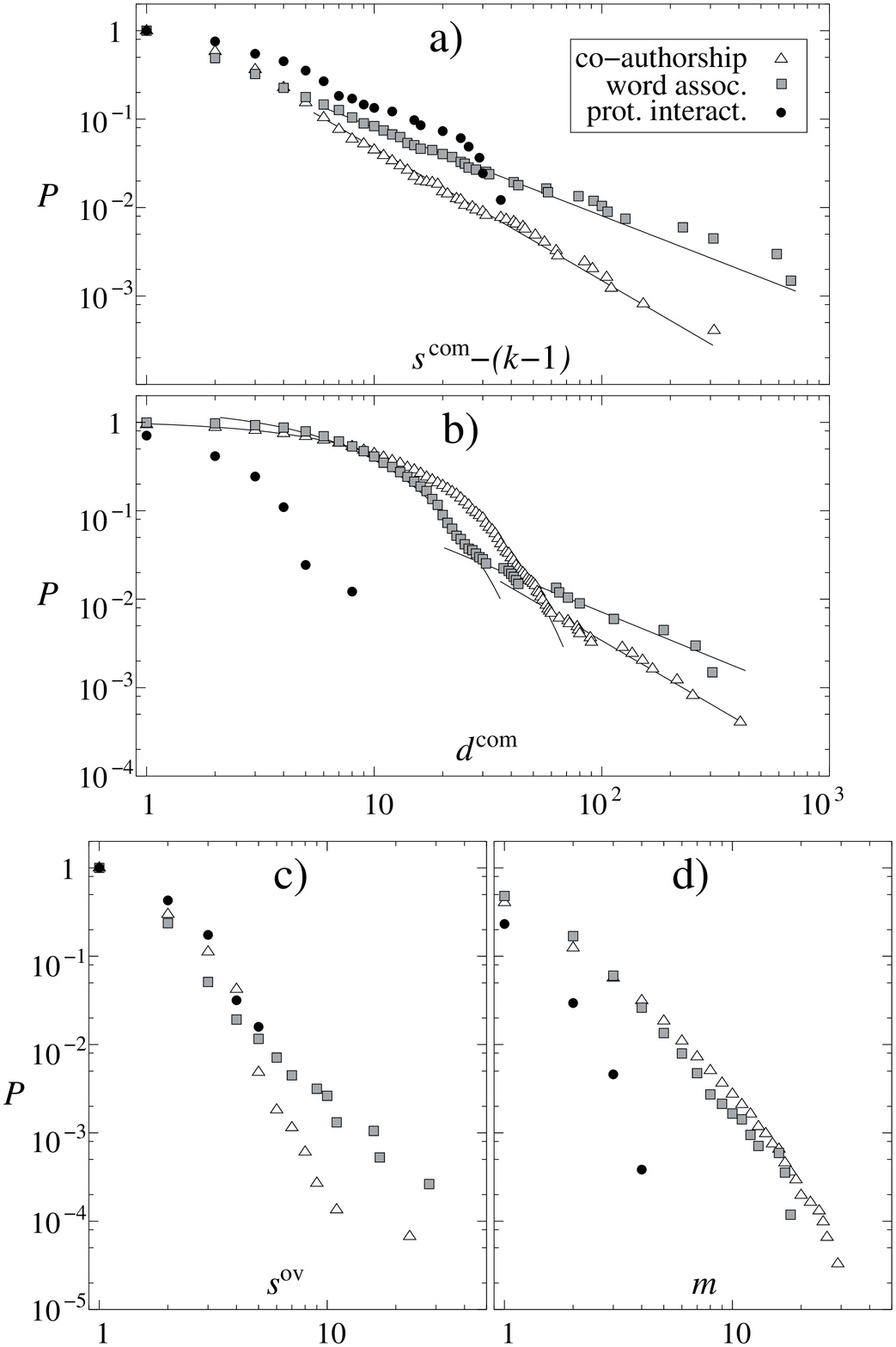}}
\caption{
Statistics of the $k$-clique-communities for three large networks.
These are the co-authorship network
of the Los Alamos cond-mat archive (triangles, $k$=6, $f^*=0.93$), the
word association network of the South Florida Free Association norms
(squares, $k=4$, $f^*=0.67$), and the protein interaction network of
the yeast \textit{S. cerevisiae} from the DIP database (circles, $k=4$).
(a) The cumulative distribution function of the community size follows a power
law with exponents between $-1$ (upper line) and $-1.6$ (lower line).
(b) The cumulative distribution of the community degree starts
exponentially and then crosses over to a power law (with the same
exponent as for the community size distribution).
Plot (c) is the cumulative distribution of the overlap size and
(d) is that of the membership number.
}
\label{fig:stats}
\end{figure}

The four distributions characterising the global community structure of
these networks are displayed in Fig.~\ref{fig:stats}.
Although the scaling of the size of non-overlapping communities has
already been demonstrated for social networks
\cite{newman-pre,radicchi-pnas},
it is striking to observe how this aspect of large real networks is
preserved even when a more complete picture (allowing overlaps)
is investigated. In Fig.~\ref{fig:stats}a the power law
dependence $P(\comsize)\sim(\comsize)^{-\tau}$ with an exponent ranging
between $\tau=1$ and $1.6$ is well pronounced and is valid
nearly over the entire range of community sizes. 

It is well known
\cite{barabasi-albert,albert-revmod,dm-book}
that the nodes of large real networks have a power law degree
distribution. Will the same kind of distribution hold when we move
to the next level of organisation and consider the degrees of the
communities?
Remarkably, we find that it is not the case.
The community degrees (Fig.~\ref{fig:stats}b) have a
very unique distribution, consisting of two distinct parts: an
exponential decay $P(\comdeg)\sim\exp(-\comdeg/\comdeg_0)$ with a
characteristic community degree $\comdeg_0$
(which is in the order of $\left<\comdeg\right>$ shown in
Table~\ref{table:com_graph}), followed by a power law
tail $\sim(\comdeg)^{-\tau}$. This new kind of behaviour is consistent
with the community size distribution assuming that on average each
node of a community has a contribution $\delta$ to the community
degree. The tail of the community degree distribution is, therefore,
simply proportional to that of the community size distribution. At the
first part of $P(\comdeg)$, on the other hand,
a characteristic scale $\comdeg_0\sim k\delta$ appears, because the
majority of the communities have a size of the order of $k$ (see
Fig.~\ref{fig:stats}a) and their distribution around $\comdeg_0$
dominates this part of the curve.
%
%
Thus, the
degree to which $P(\comdeg)$ deviates from a simple scaling depends on
$k$ or, in other words, on the prescribed minimum cohesiveness of the
communities.



The extent to which different communities overlap is also a relevant
property of a network. Although the range of overlap sizes is
limited, the behaviour of the cumulative overlap size
distribution $P(\ovsize)$, shown in Fig.~\ref{fig:stats}c, is close to
a power law for each network, with a rather large exponent.
We can conclude that there is no characteristic overlap size in the
networks.  Finally, in Fig.~\ref{fig:stats}d we 
display the cumulative
distribution of the membership number, $P(m)$. These plots demonstrate
that a node may belong to a number of
communities. In the collaboration and the word association network
there seems to be no characteristic value for the membership number:
the data are close to a power law dependence, with a large
exponent. In the protein interaction network, however, the
largest membership number is only 4, which is consistent with the also
rather short distribution of its community degree. 
{\corr To show that 
the communities we find are not due to some sort of 
artifact of our method, we have also determined the above
distributions for ``randomised'' graphs with parameters 
(size, degree
sequence, $k$ and $f^*$) being the same as those of 
our three examples, but with links stochastically redistributed
among the nodes. We have 
found that indeed the distributions are extremely truncated,
signifying a complete lack of the rich community
structure determined for the original data.}

In Table~\ref{table:com_graph} we have collected a few interesting
statistical properties of the network of communities.
It should be pointed out that the average clustering coefficients
$\left< C^{\rm com}\right>$ are relatively high, indicating
that two communities overlapping with a given community are likely to
overlap with each other as well, mostly because they all
share the same overlapping region. The high fraction of shared nodes
is yet another indication of the importance of overlaps between
the communities.

\begin{table}[!t]
{\centerline{
\begin{tabular}{|r|r|r|r|r|}\hline
 & $N^{\rm com}$
 & $\left<\comdeg\right>$
 & $\left<C^{\rm com}\right>$
 & $\left<r\right>$ \cr
\hline
 co-authorship    & 2450 &  12.10  &  0.44  &  58\%  \cr \hline
 word assoc.      &  670 &  11.33  &  0.56  &  72\%  \cr \hline
 prot. interact.  &   82 &   1.54  &  0.17  &  26\%  \cr \hline
\end{tabular}
}}
\caption{
Statistical properties of the network of communities.
$N^{\rm com}$ is the number of communities,
$\left<\comdeg\right>$ is the average community degree,
$\left<C^{\rm com}\right>$ is the average clustering coefficient of the
network of communities, and
$\left<r\right>$ represents the average fraction of
shared nodes in the communities.
}
\label{table:com_graph}
\end{table}

The specific scaling of the community degree distribution 
is a novel signature of the
hierarchical nature of the systems we study. We find that if we
consider the network of communities instead of the nodes themselves, we
still observe a degree distribution with a fat tail, but a
characteristic scale appears, below which the distribution is
exponential. This is consistent with our understanding of a complex
system having different levels of organisation with units specific to
each level.
In the present case the principle of organisation
(scaling) is preserved (with some specific modifications) when going to
the next level
{\corr
in good agreement with the recent finding of the
self-similarity of many complex networks
\cite{makse-nature}.}

With recent technological advances, huge sets of data are accumulating
at a tremendous pace in various fields of human activity (including
telecommunication, the internet, stock markets) and
in many areas of life and social sciences
(biomolecular assays, genetic maps, groups of web users, \etc.).
Understanding both the universal and specific features of the networks
associated with these data has become an actual and important task.
The knowledge of the community structure enables
the prediction of some essential features of the systems under
investigation. For example, since with our approach it is possible to
``zoom'' onto a single unit in a network and uncover its communities
(and the communities connected to these, and so on), we provide a tool
to interpret the local organisation of large networks and can predict
how the modular structure of the network changes if a unit is removed
(\eg, in a gene knock out experiment).
A unique feature of our method is that we can simultaneously
look at the network at a
higher level of organisation and locate the communities that play a key
role within the web of communities. Among the many possible
applications is a more sophisticated approach to the spreading of
infections (\eg, real or computer viruses) or information
in highly modular complex systems.


\bigskip
\textbf{Supplementary Information} accompanies the paper on
\textbf{www.nature.com/nature}.

\bigskip
\textbf{Acknowledgements}
 We thank A.-L. Barab\'asi and P. Pollner for
 useful discussions. We acknowledge the help of B. Kov\'acs and
 G. Szab\'o in connection with visualisation and software support.
 This research was supported by the Hungarian Research Grant
 Foundation (OTKA).

\bigskip
\textbf{Competing interests statement}
 The authors declare that they have no competing financial interests.

\bigskip
\textbf{Correspondence} and requests for materials
 should be addressed to T.V. (vicsek@angel.elte.hu).

\end{document}